\documentclass[useAMS,usenatbib]{mn2e}

\usepackage{graphicx}
\usepackage{amsmath}
\usepackage{amsfonts}
\usepackage[usenames,dvipsnaes]{xcolor}
\usepackage{dsfont}
\usepackage{color}
\usepackage{url}
\usepackage{times,amssymb,epstopdf}

\newcommand\tcr{\textcolor{black}}
\newcommand\tcb{\textcolor{black}}
\newcommand\tcm{\textcolor{black}}
\newcommand\tcrr{\textcolor{black}}
\newcommand\mk{\textcolor{black}}
\newcommand\fb{\textcolor{black}}
\newcommand\fbtwo{\textcolor{black}}

\begin{document}
 
\title[ISW in $f(R)$]{The Integrated Sachs-Wolfe effect in $f(R)$ gravity}
\author[Cai et al.]{Yan-Chuan Cai$^{1}$\thanks{E-mail: y.c.cai@durham.ac.uk}, Baojiu Li$^{1}$,  Shaun Cole$^{1}$, Carlos S. Frenk$^{1}$ and Mark Neyrinck$^{2}$ \\
 $^1$Institute for Computational Cosmology, Department of Physics, University of Durham, South Road, Durham DH1 3LE, UK \\
$^2$Department of Physics and Astronomy, The Johns Hopkins University, 3701 San Martin Drive, Baltimore, MD 21218, USA}
\maketitle
 
\begin{abstract}
\fb{We study the late-time Integrated Sachs-Wolfe (ISW) effect in $f(R)$ gravity using N-body simulations. In the $f(R)$  model under study, the linear growth rate is larger than that in general
relativity (GR). This slows down the decay of the cosmic potential and
induces a smaller ISW effect on large scales. Therefore, the $\dot\Phi$ (time
derivative of the potential) power spectrum at $k<0.1h$/Mpc is suppressed relative to 
that in GR. In the non-linear regime, relatively rapid structure formation in $f(R)$ gravity boosts the
non-linear ISW effect relative to GR, and the $\dot\Phi$ power spectrum at $k>0.1h$/Mpc is
increased (100$\%$ greater on small scales at $z=0$). We explore the
detectability of the ISW signal via stacking supercluster and
supervoids. The differences in the corresponding ISW cold or hot spots
are $\sim 20\%$ for structures of $\sim 100$Mpc/$h$. Such differences are greater for
smaller structures, but the amplitude of the signal is lower. 
The high amplitude of ISW signal detected by Granett et al.
can not explained in the $f(R)$ model. We find relatively big differences between $f(R)$ and GR in the transverse bulk motion of matter, and discuss its detectability
via the relative frequency shifts of photons from multiple lensed images.}
\end{abstract}

\section{introduction}

Models of modified gravity (MG) are introduced to explain the observed accelerating cosmic expansion \citep{Clifton:2011jh}, 
without invoking a cosmological constant in the Einstein equation. \tcrr{Scalar-tensor gravity is among them}. In
these theories, the scalar field is coupled to matter or curvature, which leads to an universal
enhancement of gravity (commonly known as the fifth force). The enhanced gravity violates existing robust tests of general relativity 
(GR) in the solar system, 
so that only theories with some {\it screening mechanism} \citep{Khoury:2010xi}
to suppress the fifth force in high density
regions are observationally viable. Gravity is therefore back to GR in
the early universe, as well as in the vicinity of virialized
objects where the local density is sufficiently high. MG models like $f(R)$ gravity with 
the chameleon screening mechanism \citep{Khoury2004} may therefore pass current solar 
system tests \citep{Will:2010xi}. Nevertheless, structure formation in these models should
be somewhat different from that of the standard $\Lambda$-cold-dark-matter ($\Lambda$CDM, where $\Lambda$ represents 
the cosmological constant) paradigm, especially in the quasi-linear and nonlinear regime. 

Previous work with the chameleon type of model has shown that there is an increase of the abundance of massive halos and voids 
compared to a GR $\Lambda$CDM universe \citep{Schmidt:2008tn,Zhao:2010qy,Li:2011pj}, due to the enhanced gravity near clusters and inside voids \citep{LiEfstathiou2012, Clampitt2013}.
An equivalent view to this is that massive halos are becoming heavier and voids larger, with the effect more significant in voids  
than in halos \citep{Li:2011pj}. Moreover, the expansion of voids is also faster \citep{Clampitt2013}. 
While voids occupy the majority of the volume of the Universe, it is difficult 
to probe into this emptiness due to the lack of tracer. For this reason, the Integrated Sachs Wolfe (ISW) effect \citep{Sachs1967} 
is an important probe. 

The ISW effect arises from the time variation of the cosmic gravitational potential. In the linear regime of a $\Lambda$CDM-like universe,
late time cosmic accelerations causes the potential to decay. CMB photons gain (lose) 
energy when traversing the decaying potential wells (hills). This induces ISW hot (cold) spots that are 
associated with superclusters and supervoids. Naively, the fact that superstructures grow larger and deeper seems to suggest a 
stronger ISW signal in both overdense 
and underdense regions in $f(R)$ models. This being true, it may help ease the tension between the detected ISW cold and hot spots 
with predictions of a $\Lambda$CDM model. The measured ISW signal from the stacking of
4-deg$^2$-size regions of the CMB corresponding to the SDSS superclusters and supervoids is found to be $2$-$3\sigma$ higher
than estimates from simulations \citep{Granett2008, Papai11}. This
tension with the $\Lambda$CDM paradigm is perhaps more than $3\sigma$ 
as suggested in \citet{Nadathur2012}, \citet{HM12} and \citet{Flender2013}. 
\fbtwo{See also the lastest analysis using SDSS-DR7 void catalogues \citep{Cai2013,Ilic2013, Planck}.} 
One plausible explanation 
of this discrepancy is that the size and depths of superstructures in the real Universe may 
be greater than expected in GR, which is what an $f(R)$ model suggests.

In this work, we will address this particular problem using N-body simulations of an $f(R)$ and GR $\Lambda$CDM universe. 
With the same initial conditions and the $f(R)$ model parameters tuned to have the same expansion 
history as GR, we are able to compare the differences of structure formation and evolution of 
the cosmic potentials in these two models solely due to the fifth force. We aim at exploring the physics of the 
ISW effect as well as its non-linear aspect in $f(R)$ models and address quantitatively the differences with GR 
in the ISW signal, in particular, the stacking of superstructures. We will discuss the detectability of the 
differences between models.

The outline of this paper is as follows: In section \ref{sect:theory}, we give a brief summary of the coupled scalar field
gravity, of which the $f(R)$ model is an example. Section \ref{sect: powerspectrum} shows simulations of the ISW effect in both the 
$f(R)$ model and $\Lambda$CDM. In section \ref{sect:coldspot}, we explore the stacking of the ISW signal for superclusters and supervoids in these two models, 
and present an example of strong non-linear structure in section \ref{sect: movingcluster}. We summarize our results and discuss detectabilities 
in section \ref{sect: conclusion}.

\section{Introduction to $f(R)$ gravity}

\label{sect:theory}

In this section we briefly describe the essentials of the theory and simulations of $f(R)$ gravity.

\subsection{$f(R)$ modified gravity}

Although also considered in other contexts, the $f(R)$ gravity model has received a lot of recent attention mainly because it provides a plausible explanation to the accelerating cosmic expansion \citep{Carroll:2003wy,Carroll:2004de}. The idea is to replace the cosmological constant $\Lambda$ in the standard Einstein-Hilbert action with an algebraic function of the Ricci scalar $R$:
\begin{eqnarray}
S &=& \int{\rm d}^4x\sqrt{-g}\left[\frac{1}{2}M^2_{\rm Pl}\left(R+f(R)\right)+\mathcal{L}_{\rm m}\right],
\end{eqnarray}
in which $g$ is the determinant of the metric tensor $g_{\mu\nu}$, $M_{\rm Pl}$ is the reduced Planck mass which satisfies $M^{-2}_{\rm Pl}=8\pi G$, $G$ is Newton's constant, and $\mathcal{L}_{\rm m}$ is the Lagrangian density for matter (including baryonic and dark matter). When the curvature is sufficiently high, the function $f(R)$ approaches a constant value and behaves as a cosmological constant. In other regimes, however, the fact that $f_R\equiv{\rm d}f(R)/{\rm d}R\neq0$ brings in complicated dynamics, and gives the theory a rich phenomenology \citep{Sotiriou:2008rp,DeFelice:2010aj}.

The modified Einstein equation for $f(R)$ gravity can be derived by varying the above action with respect to the metric tensor $g_{\mu\nu}$:
\begin{eqnarray}\label{eq:einstein_eqn}
G_{\mu\nu} + f_RR_{\mu\nu} - \left[\frac{1}{2}f-\Box f_R\right]g_{\mu\nu} - \nabla_\mu\nabla_\nu f_R = 8\pi GT^{\rm m}_{\mu\nu},
\end{eqnarray}
where $\Box\equiv\nabla^\mu\nabla_\mu$ with $\nabla_\mu$ being the covariant derivative, $G_{\mu\nu}\equiv R_{\mu\nu}-\frac{1}{2}g_{\mu\nu}R$ is the Einstein tensor with $R_{\mu\nu}$ being the Ricci tensor, and $T^{\rm m}_{\mu\nu}$ is the energy momentum tensor for matter. Note that the appearance of second-order derivatives in front of $f_R$ makes the above equation fourth-order in nature, since $f_R$ is a function of $R$, which itself already contains second-order derivatives of the metric tensor. One could, however, consider $f_R$ as a new scalar degree of freedom (the {\it scalaron}), the equation of motion of which is obtained by taking the trace of Eq.~(\ref{eq:einstein_eqn}) as
\begin{eqnarray}\label{eq:scalaron_eqn}
\Box f_R &=& \frac{1}{3}\left[R-f_RR+2f(R)+8\pi G\rho_{\rm m}\right],
\end{eqnarray}
in which $\rho_{\rm m}$ is the energy density for matter. In the GR limit, $f_R\rightarrow0$, $f(R)\rightarrow 2\Lambda$ so that \fb{$R=-8\pi G\rho_{\rm m}+4\Lambda$}, which is a well-known result. In this limit, $f_R$ becomes non-dynamical and Eq.~(\ref{eq:einstein_eqn}) returns to second order.

We are therefore led to the following natural conclusion: for a $f(R)$ model to pass the stringent solar system tests, the dynamics of Eq.~(\ref{eq:scalaron_eqn}) must be such that $f_R\rightarrow0$ when \fb{$\rho_{\rm m}\rightarrow\infty$}; failing this and the model would be ruled out by local experiments. This is essentially the idea underlying the chameleon mechanism \citep{Khoury2004}. To see this more explicitly, let us define an effective potential $V_{\rm eff}(f_R; \rho_{\rm m})$ for the scalaron field, in which the dependence on $\rho_{\rm m}$ is made explicit, through
\begin{eqnarray}
\frac{{\rm d}V_{\rm eff}(f_R;\rho_{\rm m})}{{\rm d}f_R} &=& \frac{1}{3}\left[R-f_RR+2f(R)+8\pi G\rho_{\rm m}\right].
\end{eqnarray}
If, for a given $f(R)$ model, one has $|f(R)|\ll|R|$ and $|f_R|\ll1$ where the matter density is high, then $R=-8\pi G\rho_{\rm m}$ (the GR solution) extremises $V_{\rm eff}(f_R;\rho_{\rm m})$ and it is a minimum in $V_{\rm eff}$ if
\begin{eqnarray}
m^2_{\rm eff}\ \equiv\ \frac{{\rm d}^2V_{\rm eff}(f_R;\rho_{\rm m})}{{\rm d}f_R^2}\ \approx\ \frac{1}{3}\frac{{\rm d}R}{{\rm d}f_R}\ \equiv\ \frac{1}{3}\frac{1}{f_{RR}}\ >\ 0.
\end{eqnarray}
Here, $m_{\rm eff}$ is the effective mass of the scalaron field $f_R$. If $f(R)$ varies extremely slowly with $R$ when $|R|$ is large, then it is easy to have $f^{-1}_{RR}\gg H^2$, where $H$ is the Hubble expansion rate: this is the familiar property of a chameleon field, that in  high density regions the Yukawa force mediated by the scalaron, which is proportional to $\exp(-m_{\rm eff}r)$, decays rapidly over a distance of $r\sim\mathcal{O}(1)~m^{-1}_{\rm eff}$ from the source, such that any deviation from GR is strongly suppressed.

The requirement of passing solar system tests does not place {\it direct} constraints on the properties of the function $f(R)$ in the low-curvature regime ($|R|\sim H_0^2$, where $H_0$ is the present-day Hubble expansion rate). Indeed, in these regions $m^{-1}_{\rm eff}\sim\mathcal{O}(1-10)$ Mpc in many $f(R)$ models\footnote{It cannot be drastically larger than this if the model is to pass solar system tests \citep{Brax:2012gr}}, which means that gravity can be modified by the Yukawa force mediated by the scalaron field on such scales. To see how large this modification can be, we shall first write down the perturbation equations for the scalaron field and the Newtonian potential. We work in the Newtonian gauge for which the metric element is given by
\begin{eqnarray}
{\rm d}s^2 &=& a^2(\eta)\left[(1+2\phi){\rm d}\eta^2-(1-2\psi){\rm d}x^i{\rm d}x_i\right],
\end{eqnarray}
in which $\phi(\eta,{\bf x})$ and $\psi(\eta,{\bf x})$ are respectively the Newtonian potential and the perturbation to the spatial curvature, and are functions of the conformal time $\eta$ and comoving coordinate ${\bf x}$. Here $a$ is the scale factor which is normalised to $1$ today. On scales much smaller than the Hubble scale $H^{-1}_0$, we can follow the quasi-static approximation and neglect the time derivatives of small quantities such as $f_R$ and $\Phi$ \citep{Oyaizu:2008sr}. Then the scalaron equation of motion can be written as
\begin{eqnarray}\label{eq:scalaron_eqn_sim}
\vec{\nabla}^2f_R &=& -\frac{1}{3}a^2\left[R\left(f_R\right)-\bar{R}+8\pi G\left(\rho_{\rm m}-\bar{\rho}_{\rm m}\right)\right],
\end{eqnarray}
in which $\vec{\nabla}$ denotes the spatial derivative and $R$ has been explicitly written as a function of $f_R$. Similarly, the modified Poisson equation simplifies to
\begin{eqnarray}\label{eq:poisson_eqn}
\vec{\nabla}^2\phi &=& \frac{16\pi G}{3}a^2\left(\rho_{\rm m}-\bar{\rho}_{\rm m}\right) + \frac{1}{6}a^2\left[R(f_R)-\bar{R}\right].
\end{eqnarray}
In the low-curvature regime, the curvature scalar $R$ changes slowly with $\rho_{\rm m}$ (this is different from the high-curvature regime discussed above, where we have $R\propto\rho_{\rm m}$), and the $R-\bar{R}$ term in Eq.~(\ref{eq:scalaron_eqn_sim}) is relatively unimportant, which leads to 
\begin{eqnarray}
\vec{\nabla}^2f_R &\approx& -\frac{8}{3}\pi Ga^2\left(\rho_{\rm m}-\bar{\rho}_{\rm m}\right),
\end{eqnarray}
and similarly from Eq.~(\ref{eq:poisson_eqn}) we find
\begin{eqnarray}
\vec{\nabla}^2\phi &\approx& 4\pi G_{\rm eff}a^2\left(\rho_{\rm m}-\bar{\rho}_{\rm m}\right),
\end{eqnarray}
in which we have defined $G_{\rm eff}=4G/3$. This implies a factor-of-$4/3$ enhancement of the strength of gravity compared to GR, which is the largest deviation from GR that is allowed by any $f(R)$ model, regardless of the exact functional form of $f(R)$.

Therefore, a successful $f(R)$ model is one which reduces to GR ($f_R\rightarrow0$) in high-density regions by virtue of the chameleon mechanism, while possibly allowing a $4/3$ enhancement of gravity in low-density regions (below $m^{-1}_{\rm eff}$). The deviation from GR diminishes again beyond $m^{-1}_{\rm eff}$, which is typically of order Mpc, and this implies that the largest scales, such as those probed by the CMB, are not affected. Such a scale dependence of MG effects in $f(R)$ gravity is very well known \citep{Song:2006ej,Li:2007xn} and gives rise to a scale-dependent linear growth factor.

\subsection{The Hu-Sawicki model}

We have seen above that there are scale and environmental dependences of the effect of $f(R)$ gravity, both determined by the quantity $m^{-2}_{\rm eff}\approx3f_{RR}$. Therefore, the model behaviour is completely specified if and only if the functional form of $f(R)$ is known. Some choices of $f(R)$ can be found in the early works of \citet{Faulkner:2006ub,Navarro:2006mw,Li:2007xn} and \citet{Brax:2008hh}. The fourth-order nature of $f(R)$ gravity even enables us to make the background expansion history exactly mimic that of the $\Lambda$CDM paradigm \citep{Song:2006ej}.

In this work we choose to work with an $f(R)$ model first proposed by \citet{Hu:2007nk}, for which
\begin{eqnarray}\label{eq:hs_fr}
f(R) &=& -M^2\frac{c_1\left(-R/M^2\right)^n}{c_2\left(-R/M^2\right)^n+1},
\end{eqnarray}
where $M$ is a characteristic mass scale and $M^2=8\pi G\bar{\rho}_{\rm m0}/3=H^2_0\Omega_{\rm m}$, with $\bar{\rho}_{\rm m0}$ and $\Omega_{\rm m}$ being the present-day background density and fractional energy density for matter respectively.  $c_1,c_2,n$ are dimensionless parameters. Note that, as the current value of $\bar{R}$, assuming a $\Lambda$CDM background expansion history with $\Omega_{\rm m}=0.24$ and $\Omega_{\Lambda}\equiv1-\Omega_{\rm m}=0.76$, is $|\bar{R}|\approx41M^2\gg M^2$, $f(R)$ is essentially constant throughout the cosmic evolution in this model\footnote{This can be seen by substituting $|R|\gg M^2$ into Eq.~(\ref{eq:hs_fr}). Recall that $|\bar{R}|$ is even larger at earlier times so that this conclusion holds at all times.}, satisfying the two requirements that $f(R)$ behaves as a cosmological constant in the background cosmology and that it varies extremely slowly with $R$ (especially when $|R|$ is large) to guarantee a working chameleon mechanism. 

As $\bar{m}_{\rm eff}\sim\bar{f}^{-1/2}_{RR} \gg H$ (as can be easily checked) throughout the cosmic history, the scalaron stays at the minimum of its effective potential $V_{\rm eff}(f_R;\rho_{\rm m})$, where it oscillates quickly \citep{Brax:2012gr}. This minimum is given by setting ${\rm d}V_{\rm eff}/{\rm d}f_R=0$ as
\begin{eqnarray}
-\bar{R}\ \approx\  8\pi G\bar{\rho}_{\rm m}+2\bar{f}\ \approx\ 3M^2\left(a^{-3}+\frac{2c_1}{3c_2}\right),
\end{eqnarray}
in which we have used $|f_RR|\ll|R|$ in the first equality and the asymptotic form of $f(R)$, Eq.~(\ref{eq:hs_fr}), in the second equality. To match the $\Lambda$CDM background evolution, we set 
\begin{eqnarray}
\frac{c_1}{c_2} &=& 6\frac{\Omega_\Lambda}{\Omega_{\rm m}}.
\end{eqnarray}
This reduces the number of independent parameters for the Hu-Sawicki model ($c_1,c_2,n$) to two.

In the limit $|R|\gg M^2$, we find
\begin{eqnarray}
f_R &\approx& -n\frac{c_1}{c_2^2}\left(\frac{M^2}{-R}\right)^{1+n}.
\end{eqnarray}
In particular, the value of the scalaron today, $f_{R0}$, satisfies
\begin{eqnarray}
\frac{c_1}{c_2^2} &=& -\frac{1}{n}\left[3\left(1+4\frac{\Omega_\Lambda}{\Omega_{\rm m}}\right)\right]^{1+n}f_{R0}.
\end{eqnarray}
Therefore, the model can be conveniently specified by two parameters: $n$ and $f_{R0}$. If $n$ is fixed, the smaller $|f_{R0}|$ is, the stronger the chameleon effect is, because it is more difficult for $|f_R|$ to become large anywhere. In the special case of $n=1$, cluster abundance data places a constraint $|f_{R0}|\lesssim10^{-4}$ \citep{Schmidt:2009am}, and studies of other observables point to similar constraints \citep{Jennings2012,Hellwing:2013rxa}.

\subsection{Numerical simulations}

The fact that the behaviour of $f(R)$ gravity depends on its local environmental makes the model highly nonlinear. Linear perturbation theory works reliably only for the largest scales (e.g., in predicting the CMB power spectrum), at which gravity is essentially GR and it fails whenever there are significant deviations from GR \citep{Li:2012by}. This highlights the necessity of nonlinear numerical simulations in studies of large-scale structure formation in MG theories \citep{Oyaizu:2008sr,Schmidt:2008tn,Li:2009sy,Li:2010mqa,Zhao:2010qy,Li:2010re}.

Our study in this paper is based on N-body simulations carried out using the {\sc ecosmog} code \citep{Li:2011vk}, which is an extension of the publicly available {\sc ramses} code \citep{Teyssier:2001cp} to do generic MG simulations. The simulations used here are the same as those used in previous studies of redshift-space distortions \citep{Jennings2012} and high-order hierarchical clustering \citep{Hellwing:2013rxa}, and we will only very briefly describe them; more details can be found in those papers.

The models simulated are the Hu-Sawicki model described in the above subsection, with $n=1$ and $-f_{R0}=10^{-6}, 10^{-5}, 10^{-4}$ respectively. \mk{Larger values of $|f_{R0}|$ go against cosmological constraints, 
e.g, \citep{Ferraro2011} and smaller values would mean that deviations from $\Lambda$CDM are too small to show any interesting effects on the ISW. Indeed, 
our results show that even for $|f_{R0}|=10^{-6}$ and $10^{-5}$ the deviation is small.} Therefore, in what follows we only present the results for the model with $-f_{R0}=10^{-4}$. The physical parameters of the model are chosen as 
\begin{eqnarray}
&&\left\{\Omega_{\rm m}, \Omega_{\Lambda}, n_{\rm s}, h\equiv H_0/(100{\rm km/s/Mpc}), \sigma_8 \right\}\nonumber\\
&=& \left\{0.24, 0.76, 0.961, 0.73, 0.80\right\},
\end{eqnarray}
in which $n_{\rm s}, \sigma_8$ are respectively the index of the primordial power spectrum and the rms density perturbation at the present predicted by linear theory within a top-hat smoothing window of comoving size $8h^{-1}$Mpc, and are only used when setting the initial conditions for the simulations. Because the chameleon effect is so strong that our $f(R)$ model behaves indistinguishably from GR at redshift $z>10$, the initial conditions (which is set at $z_{\rm ini}=49.0$) are the same for our GR and $f(R)$ simulation, and are normalised such that $\sigma_8=0.8$ today. We used the publicly available code {\sc mpgrafic} \citep{Prunet:2008fv} to set the initial condition in parallel. As the same initial conditions are used for GR and $f(R)$ simulations, they have the same initial random phases, which makes the model comparison easier (for example, in most cases we can find the `same' dark matter halo or void in both simulations).

We have run six realisations of GR and $f(R)$ simulations to beat down the statistical error. The simulations were done in parallel with {\sc mpi}, using 504 processors, on the {\sc cosma} supercomputer hosted by the Institute for Computation Cosmology at Durham University. The box size is $B=1.5$Gpc$/h$ and $1024^3$ particles are used in the simulations. The evolution starts on a regular mesh with $1024^3$ cubic cells covering the whole computational domain, which then adaptively self-refine when the particle number inside a cell exceeds 8 to achieve high force resolution in dense regions (which reaches $\sim22.9$kpc$/h$ on the finest level).

The halo catalogues were identified using the spherical overdensity algorithm implemented in the {\sc ahf} code \citep{Knollmann:2009pb}.

\begin{figure}
\advance\leftskip -0.3cm
\scalebox{0.45}{
\includegraphics[angle=0]{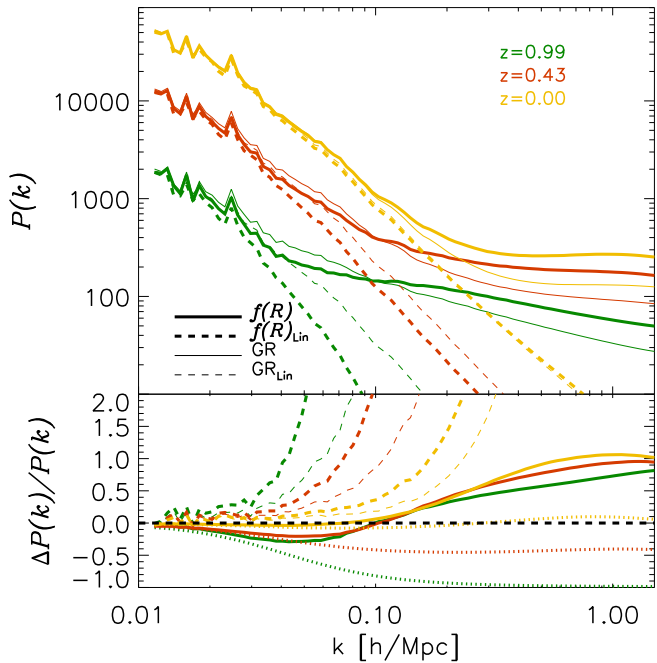}}
\caption{Comparing power spectra of $\dot\Phi$ between $f(R)$ (thick lines) and GR (thin lines) simulations at 
different redshifts. The power spectra have been rescaled as 
${\mathcal P}(k)=P_{\dot\Phi\dot\Phi}(k)/[\frac{3}{2}\left(\frac{H_0}{k}\right)^2\Omega_{\rm m}\dot a/a^2 ]^2$.
The power spectra given by the linear theory approximation are plotted as dashed lines. 
Bottom panel shows the fractional differences of power spectra: solid lines -- 
ISWRS power spectra in $f(R)$ versus GR; thick dashed lines -- ISWRS power spectra versus linear ISW power spectra in $f(R)$; 
thin dashed lines -- ISWRS power spectra versus linear ISW power spectra in GR; dotted lines -- 
linear ISW power spectra in $f(R)$ versus that in GR.}
\label{PowerSpectrumF4}
\end{figure} 

\section{power spectrum}
\label{sect: powerspectrum}

\tcr{The ISW effect is determined by the time variation of the lensing potential $\Phi\equiv(\phi+\psi)/2$. In GR, we have $\Phi=\phi=\psi$ when the anisotropic stress from radiation can be neglected at late times. For $f(R)$ gravity, however, $\phi\neq\psi$ in general and $\phi$ satisfies a modified Poisson equation as we have seen above. Fortunately, $\Phi$ still satisfies the usual Poisson equation}
\begin{eqnarray}\label{eq:standard_poisson}
\tcr{\vec{\nabla}^2\Phi\ =\ 4\pi G(\rho_{\rm m}-\bar{\rho}_{\rm m}),}
\end{eqnarray}
\tcr{even in $f(R)$ gravity. Furthermore, the continuity and Euler equations are unchanged in $f(R)$ gravity. Consequently, the techniques developed in \citet{Cai2010} work for $f(R)$ gravity as well.}

With all the dark matter particles in each simulation box, we follow \citet{Cai2010} \fb{[see also \citep{Cai2009, Smith2009, Watson2013}]} to compute the time derivative of the potential $\dot\Phi$ using particle positions and velocities. 
This can be achieved in Fourier space using
\begin{equation}\label{eq1}
\dot{\Phi}(\vec{k},t)=\frac{3}{2}\left(\frac{H_0}{k}\right)^2\Omega_{\rm m}
\left[\frac{\dot{a}}{a^2}\delta(\vec{k},t)+\frac{i\vec{k}\cdot\vec{p}(\vec{k},t)}{a}\right],
\end{equation}
where $\vec p(\vec k,t)$ is the Fourier transform of the momentum density \tcb{divided by the mean mass density}, 
$\vec p(\vec x,t)=[1+\delta(\vec x,t)]\vec v(\vec x,t)$, and  $\delta(\vec k, t)$ is the Fourier transform of the density contrast \tcr{$\delta(\vec{x},t)$}. 
The inverse Fourier transform of the above 
yields $\dot\Phi$ in real space on 3D grids.
The integration of $\dot\Phi$ along the line-of-sight 
\begin{eqnarray}
 \fb{\frac{\Delta T(\hat n)}{T}} &=& \frac{2}{c^2} \int \dot \Phi(\hat n, t){\rm d}t,
 \end{eqnarray} 
gives the ISW temperature fluctuation including its non-linear component, the Rees-Sciama 
effect (RS) \citep{Rees68}. 
The above calculations hold true for both GR and $f(R)$ simulations. Unlike dark matter particles, photons do not feel the \tcr{modified gravitational} force in $f(R)$. 
Therefore, the difference in $f(R)$ gravity in terms of $\dot\Phi$ only arises from differences in structure formation \tcr{[cf.~Eq.~(\ref{eq:standard_poisson})]}, which is fully captured by 
Eq.~(\ref{eq1}).  

In the linear regime, the velocity field is related to the density field by the linearised continuity equation 
\fb{$\vec{p}(\vec{k},t)=i{\dot\delta(k,t)\vec k}/{k^2}\approx i(\dot a/a){\tcr{f(k,t)}\delta(k,t)\vec k}/{k^2}$}. Thus
\begin{eqnarray}\label{eq3}
\dot{\Phi}(\vec{k},t) &=& \frac{3}{2}\left(\frac{H_0}{k}\right)^2\Omega_{\rm m}
\frac{\dot{a}}{a^2}\delta(\vec k,t)[1-\tcr{f(k,t)}],
\end{eqnarray}
where \tcr{$f(k,t)$} is the linear growth rate
\tcr{$f(k,t)\equiv{{\rm d}\ln D}/{{\rm d}\ln a}$ with $D$ being the linear growth factor}.
This equation is the conventional way of modelling the ISW effect and uses only 
information from the density field. Integration of the above equation along each line of sight gives a 2D ISW 
temperature map. \tcm{Note that in GR, the linear growth rate $f$ is a constant at a given redshift, 
but in $f(R)$, it can be scale dependent} (see Sect.~\ref{sect:theory}). 
In the model chosen, $f$ starts to deviate from that of GR at $k\sim 0.01h$/Mpc, and reaches $\sim$20$\%$ at $k\sim 1h$/Mpc
\citep{Jennings2012}.

For convenience, we will call results from applying the linear approximation for the velocity field 
(Eq.$~$\ref{eq3}) the linear ISW effect, and the full calculations using Eq.~(\ref{eq1}) that include the nonlinear RS effect the ISWRS.

The power spectra of $\dot\Phi$ can be calculated using Eqs.~(\ref{eq1}, \ref{eq3}). We normalize them by the 
common factors in front of the square brackets in Eq.~(\ref{eq1}), 
${\mathcal P}(k)=P_{\dot\Phi\dot\Phi}(k)/\left[\frac{3}{2}\left(\frac{H_0}{k}\right)^2\Omega_{\rm m}\dot a/a^2  \right]^2$. Results comparing 
$f(R)$ and GR power spectra are shown in Fig. \ref{PowerSpectrumF4}. On large scales, i.e. $k<0.1h$/Mpc, the amplitudes of the $\dot\Phi$ 
power spectra can be smaller in $f(R)$ than in GR, by 20\% to 30\%. This can be understood by the following reasoning.
In $f(R)$, gravity is relatively strong at late times. It helps accelerate structure formation, driving stronger convergent 
flows of dark matter towards potential minima and makes them deeper. This counters the effect of cosmic acceleration, 
which makes cosmic potential wells shallower. On large scales, the tendency to grow of potential wells in $f(R)$ models is sub-dominant 
to their reduction due to comic acceleration, thus the net effect is slowing down the decay of potentials on large scales. 
The linear growth rate $f$ is greater in the (quasi-) linear regime, which makes the 
factor of $(1-f)$ smaller, and hence results in a smaller amplitude of the $\dot\Phi$ power spectrum in that regime. 

We also notice that in the range of redshifts shown in the figure, the fractional differences of the linear ISW $\dot\Phi$ increase with redshift (dotted lines). 
The trend is similar for the solid lines (full ISWRS) but weaker. This is more subtle to understand. The fractional difference of the $\dot\Phi$ power spectra 
is proportional to $[(1-f_{\rm GR}\Delta f)/(1-f_{\rm GR})]^2-1$, where $f_{\rm GR}$ is the GR linear growth rate and \fb{$\Delta f$ is the ratio of the growth rates between $f(R)$ and GR}, 
which can be linear or non-linear and has $k$-dependence. \tcb{In the linear regime, $\Delta f$ varies from a few percents to more than 10 percents at 
$k\sim0.01$ to $0.1h$/Mpc}, and its redshift dependence is not as strong as that of $f_{\rm GR}$. At high-$z$, $f_{\rm GR}$ is very close to unity, the denominator $(1-f_{\rm GR})$ 
can easily amplify the fractional difference of $\dot\Phi$. This explains the behaviour of the dotted lines, that the fractional difference of the linear 
$\dot\Phi$ is greater at high-$z$.  When considering the full ISWRS effect, since non-linearity is relatively more important at high-$z$ and it counters 
the linear effect in the $f(R)$ model, this makes the overall factor of $[(1-f_{\rm GR}\Delta f)/(1-f_{\rm GR})]^2-1$ smaller and the redshift evolution at 
$k\sim0.01$ to $0.1h$/Mpc weaker. This suggests that it is very important in $f(R)$ models to have the full ISWRS calculation rather than making 
the linear approximation.

At small scales, i.e. $k>0.1h$/Mpc, the growth of potential wells in $f(R)$ models is much greater and overcomes \tcrr{the tendency for them to be reduced due to the cosmic acceleration}. 
The sign of $(1-f)$ is reversed, and its magnitude increases with the strength of non-linearity.
Therefore, the amplitudes of the $\dot\Phi$ power spectra become greater in $f(R)$ gravity than in GR, \tcr{by a factor of $\sim$2 for ISWRS (solid lines)}. This difference is
about twice as much as that in the matter power spectrum \citep{Zhao:2010qy,Li:2012by}. Note that the magnitude 
of the difference is close to \tcr{$(G_{\rm eff}/G)^2-1=(4/3)^2-1$}, where $G_{\rm eff}$ is \tcr{defined in Section~\ref{sect:theory}}. This may not be surprising since the velocity field is dominant for $\dot\Phi$ in this regime \tcr{and has been affected by $G_{\rm eff}$ from quite early times due to the weak chameleon screening in the model with $-f_{R0}=10^{-4}$ (particles would travel 1/3 faster in $f(R)$ gravity than in GR if they feel $G_{\rm eff}$ for a long period)}. 

In this regime, the non-linear ISW effect in $f(R)$ gravity is just like the one in GR but 
enhanced. This is qualitatively consistent with the fact that halos and voids form earlier than in GR \citep{LiEfstathiou2012, Clampitt2013}, 
and we will demonstrate its consistency with the stacking of ISW cold and hot spots in Section \ref{sect:coldspot}.  

\tcr{At even smaller scales ($k>1h$/Mpc), the difference between the $\dot{\Phi}$ power spectra in these two models seems to decrease again. However, the resolution of our simulations is not high enough \tcrr{for us to} comment on the physics in this regime.}

\begin{figure*}
\begin{center}
\advance\leftskip 0.8cm
\vspace{-4.0 cm}



\scalebox{0.8}{
\includegraphics[angle=0]{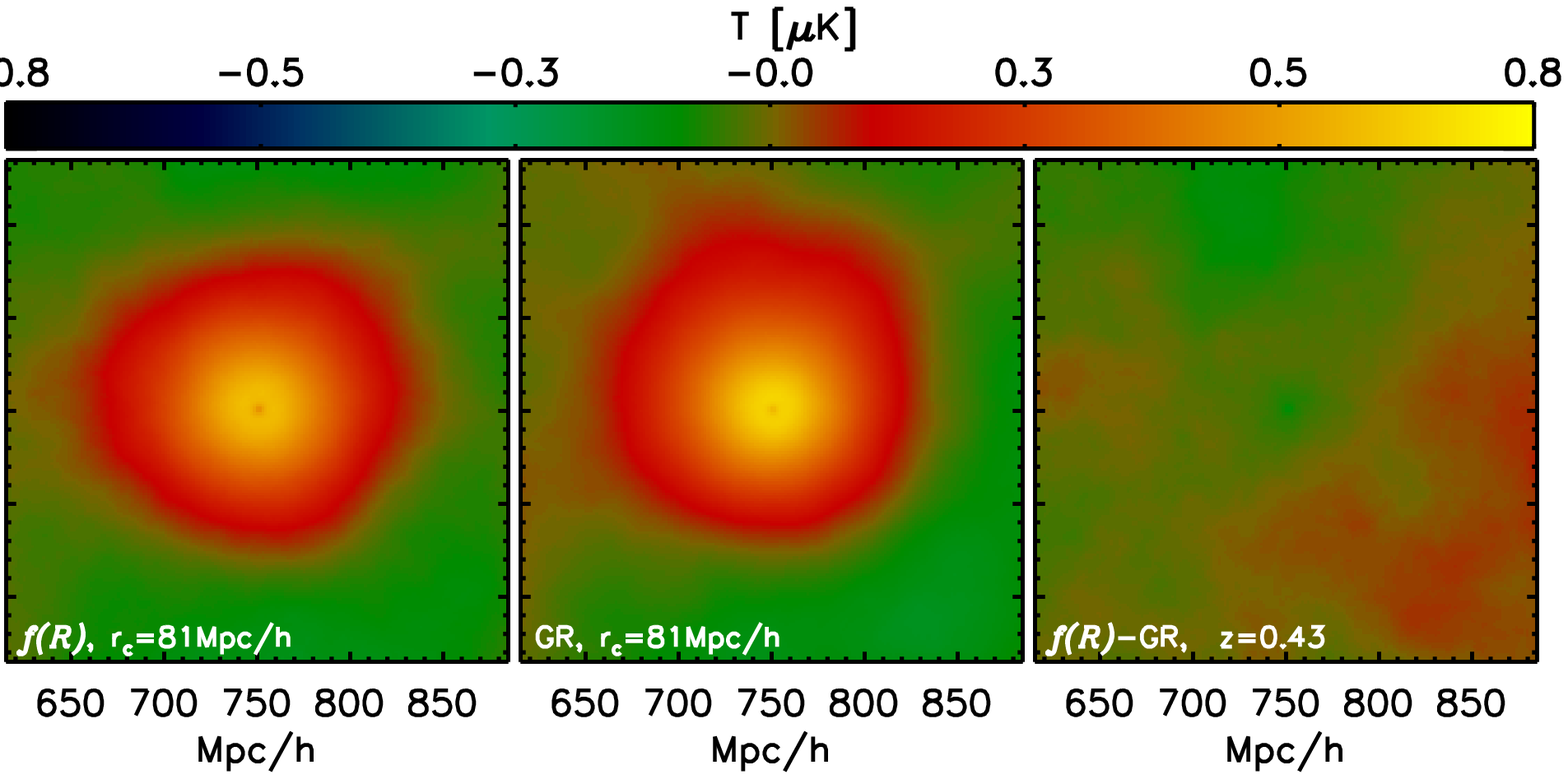}}

\vspace{-7.5 cm}

\scalebox{0.8}{
\includegraphics[angle=0]{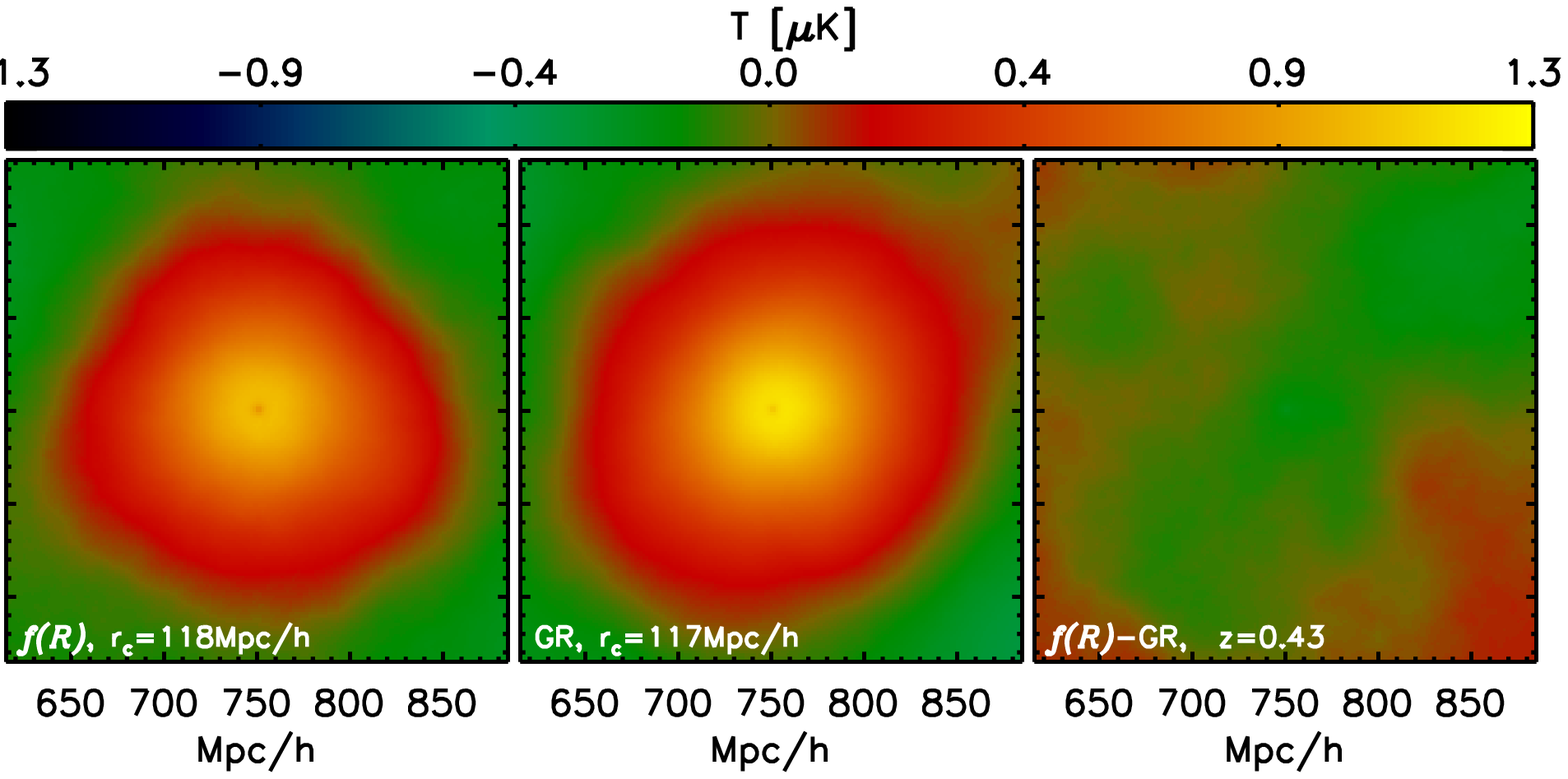}}
\vspace{-4.0 cm}
\caption{Stacks of ISWRS maps using superclusters from simulations. Superclusters of the same 
radius ranges are selected from both $f(R)$ and GR simulations, their mean radii 
are labelled in the figure. We only include structures that are 3$\sigma$ above the Poisson noise 
for the stack. Fourier $k$-modes smaller than $k=0.01 h/$Mpc are removed to reduce cosmic variance. 
The right-hand figures show the difference between $f(R)$ and GR.}
\label{StackingClusters}
\end{center}
\end{figure*}

\section{stacking of supervoids and superclusters}
\label{sect:coldspot}
\mk{Before proceeding, we pause to clearly define supervoids and superclusters. By voids and clusters, we mean objects that roughly correspond to 
outcomes of the top-hat spherical-collapse or spherical-expansion model, outlined by caustics where streams have crossed \citep{FalckEtal2012}. 
In contrast, the supervoids and superclusters we use are simply underdensities or overdensities in the halo density field, and could be much 
larger-scale than such voids and clusters. They should correspond well with such structures in the dark-matter field, in the limit of high 
sampling \citep[e.g.][]{SutterEtal2013, NeyrinckEtal2013}.}

The Poisson equation in Fourier space implies that the evolution of cosmic potential perturbations depend on the evolution of the density perturbation 
$\delta$ and the expansion history $a$, i.e. $\Phi\propto \delta/a$. In the linear regime in a GR $\Lambda$CDM universe, supervoids or 
superclusters would be stretched by the late time acceleration, $a$ grows faster than $\delta$. The reduction in the potential perturbations would 
cool down or heat up CMB photons traversing them, inducing ISW cold or hot spots. In the non-linear regime, the situations 
in supervoids and superclusters are different. $\delta$ in over-dense regions grows faster than the expansion factor. The sign of 
$\dot\Phi$ is reversed. This suppresses the linear ISW hot spot in superclusters. In contrast, at the centre of a void where it is close to empty, i.e. $\delta \sim -1$, 
it can not become any emptier. The growth of $\delta$ slows down while cosmic acceleration remains the same. Thus, the reduction in the hight of 
$\dot\Phi$ perturbation is enhanced and corresponding ISW cold spots are amplified \citep{Cai2010}.

Our $f(R)$ models can mimic the late time acceleration of the $\Lambda$CDM paradigm, and therefore similar cold or hot spots associate with superstructures 
are expected. The subtle differences of the growth history in $f(R)$ gravity from GR, as addressed in the previous section, should cause differences in the ISW 
cold and hot spots. Superstructures in $f(R)$ models may grow larger and deeper driven by the fifth force \citep{LiEfstathiou2012,Clampitt2013}.  
With our simulations, we can explore the impact of it for the ISW cold and hot spots as well as their implications for observations.

Halo catalogues from simulations with the minimal halo mass of $M_{\rm min}\sim 10^{12}M_{\odot}/h$ are used to mimic how galaxies trace dark matter. 
We feed the catalogues to the public code {\sc zobov} of \citet{Neyrinck05} and \citet{Neyrinck08} 
to find superstructures. {\sc zovob} tessellates space into Voronoi cells around each halo, and applies a watershed 
algorithm \citep[e.g.][]{Platen07} on the irregular Delaunay mesh to group those cells into zones. \fbtwo{Voids are density depressions around 
minima, and clusters are density hills around maxima}. Six independent simulations 
at $z\sim0.47$ are used. This redshift is close to the range where the abnormally cold and hot 
spots are found in observations \tcm{\citep{Granett2008}}. Only superstructures at the significance level greater than 3$\sigma$ of a Poisson point sample of the 
same density are used. \fbtwo{The significance level of supervoids is quantified by the the ratio of the density on the ridge, $\rho_{\rm r}$, 
to its lowest density separating the void from a deeper void, $\rho_{\rm min}$. A higher value of $\rho_{\rm r}/\rho_{\rm min}$ means the 
supervoid is more significant. A similar definition for superclusters is adopted, but the ratio is taken between the highest density peak and 
the density on the ridge separating the cluster from a higher-density cluster. 
See Eq. 2 and Table 1. of \citep{Neyrinck08} and Eq. 1 and Table 1 of \citep{Neyrinck05} for details of the definitions.} 

\begin{figure*}
\begin{center}
\advance\leftskip 0.8cm
\vspace{-4.0 cm}


\scalebox{0.8}{
\includegraphics[angle=0]{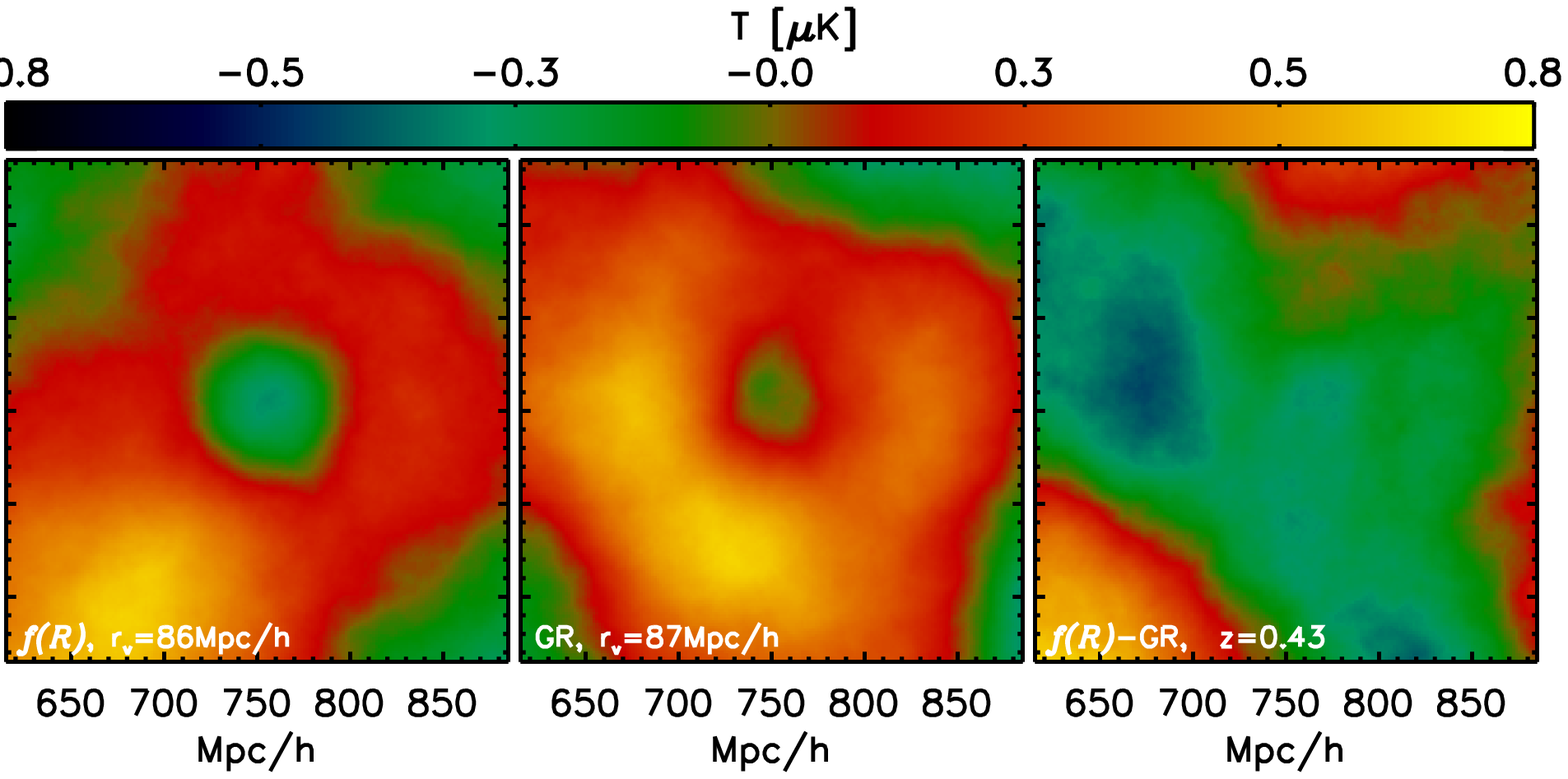}}

\vspace{-7.5 cm}

\scalebox{0.8}{
\includegraphics[angle=0]{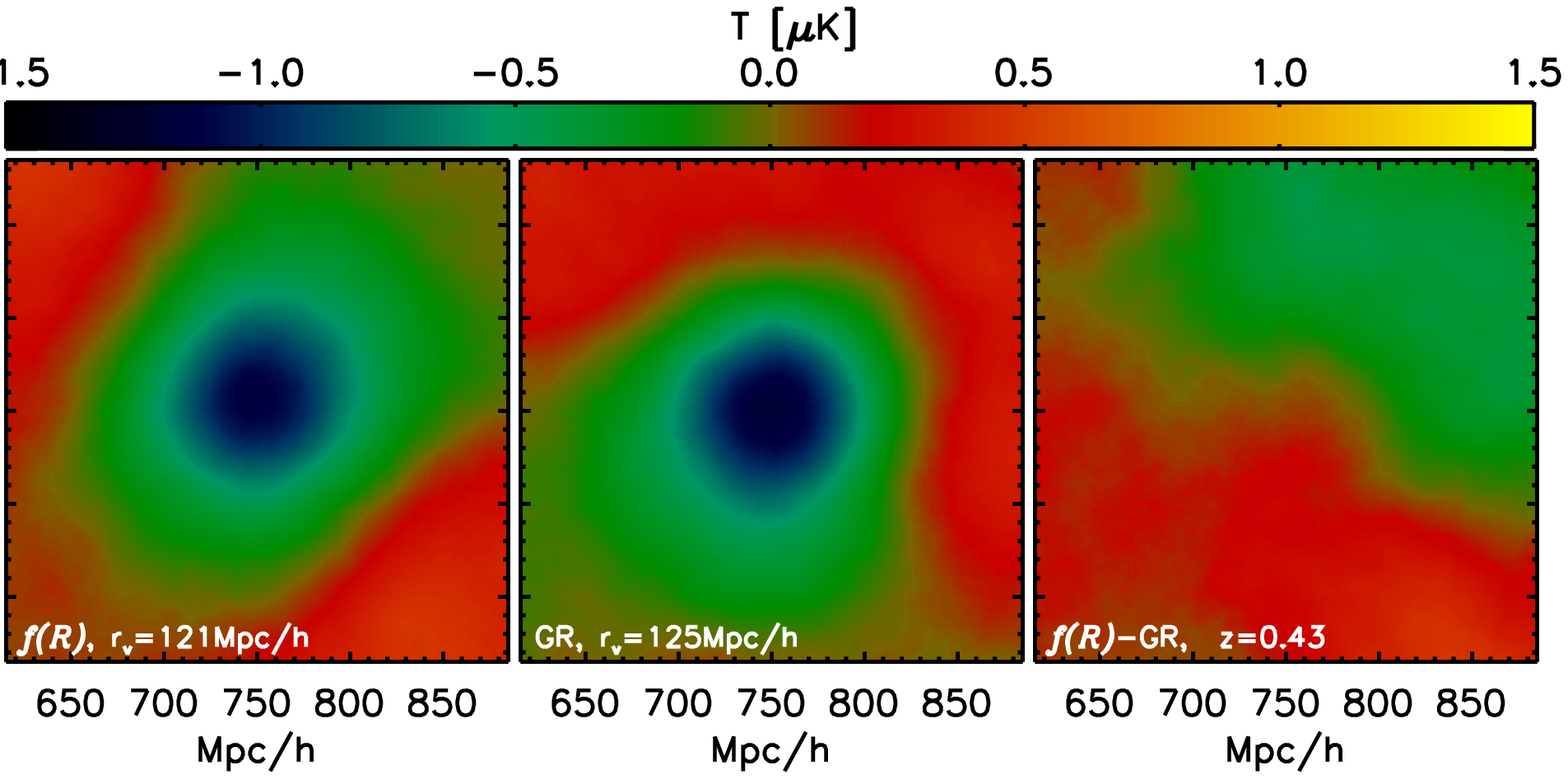}}
\vspace{-4.0 cm}
\caption{The same as Fig.$~$\ref{StackingClusters} but showing the stacking of supervoids.}
\label{StackingVoids}
\end{center}
\end{figure*}

\fbtwo{We then split the supercluster and supervoid catalogues from simulations into two different radius bins, 70-100 Mpc/$h$ and 100-150 Mpc/$h$. 
These numbers are chosen to contain enough structures such that the stacked ISW signal is able to beat cosmic variance. We make the largest radius 
bin wider to increase statistics because the largest superstructures are rarer. We will call superstructures with the radius bin of 100-150 Mpc/$h$ 
large, and those within the radial bin of 70-100 Mpc/$h$ small. The averaged radii of superstructures for each bin are labeled in the legends of 
Fig.~\ref{StackingClusters} \& \ref{StackingVoids}. In some cases, the average radii between GR and $f(R)$ differs slightly. This is expected 
because the distribution function of superstructures in these two models are different slightly. We will address even smaller structures later.} 

\tcrr{ For the simulated ISW temperature maps that we stack,} we also remove large-scale $k$-modes ($k<0.01h/$Mpc) that are much greater than 
the typical size of our superstructures. 
This is necessary to reduce the noise (cosmic variance) in the stack: due to the $1/k^2$ factor in $\dot\Phi(k)$ \tcr{[see Eq.~(\ref{eq1})]}, the ISW power spectra are 
much steeper than the matter power spectra. Thus, ISW temperature maps are always dominated by the largest Fourier $k$-mode 
of a finite simulation box, which contributes significant cosmic variance.
 
With the set up being exactly the same for the $f(R)$ and GR simulations, 
we compare results from the stacking of superstructures of similar scales 
in these two models. Fig.~\ref{StackingClusters} shows ISW hot spots corresponding to stacking 
superclusters of about 80 and 120 Mpc/$h$ in radius. It is \tcr{interesting} to find that the hot spots are 
hotter in GR than in $f(R)$, as is also evident from the temperature residuals at the centre of the hot spots 
being negative in sign. This can be understood by the fact that 
\tcm{the non-linear ISW effect at the centres of over-dense regions is countering the linear effect. The stronger non-linearity 
in $f(R)$ models makes the suppression of the linear ISW hot spot stronger, hence produces a relatively lower hot spot amplitude}, 
[see the discussion at the beginning of the section and also see \citet{Cai2010}]. In $f(R)$, as shown in the power spectrum 
of $\dot\Phi$, the non-linearity in $\dot\Phi$ is like the one in GR but enhanced. 
Therefore, the suppression to the linear ISW signal is stronger in $f(R)$, making the hot spot 
less significant than that in GR. It is also interesting to note that there is a small cold spot embedded in the 
very centre of each hot spot in both models. This is a clear sign that non-linearity at the centres of 
stacks is causing the suppression of the hot spot.

Note that for very large superclusters where the growth of structure may be in the linear regime, 
relatively lower amplitudes of hot spots in $f(R)$ gravity are also expected compared to the GR case, due to the stronger linear growth 
at $k<0.1h$/Mpc as seen in the $\dot{\Phi}$ power spectra (see Fig.~\ref{PowerSpectrumF4}). Therefore, the ISW hot spot should always be less hot in $f(R)$ models, 
whether it is in the linear or non-linear regime. 

As for supervoids, the situation is different. For the top row of Fig. \ref{StackingVoids}, where supervoids are relatively small 
($r_v\sim 80$Mpc/$h$), the ISW cold spot is colder and larger in $f(R)$, while for relatively large supervoids 
(bottom row, $r_v\sim 120$Mpc/$h$), it is less cold. It seems that whether the cold spot in $f(R)$ is colder or less 
cold than that in GR depends on the sizes of supervoids used for the stack. This again, can be understood by the 
physics of the non-linear ISW effect in supervoids. As shown before, the non-linear ISW effect in supervoids is 
opposite to that in over-dense regions, and it tends to boost its linear counterparts. Therefore, a colder and larger ISW 
cold spot in $f(R)$ than in GR is expected for supervoids of moderate sizes, because the non-linearity in $f(R)$ is stronger. 
For very large supervoids, where the growth rate is still in the linear regime, the relatively higher growth rate $f$
in $f(R)$ models makes the ISW signal [which is proportional to $(1-f)$] slightly smaller than that in GR. \fb{Therefore, 
ISW cold spot in this regime is less cold in $f(R)$ than in GR}. This is fully 
consistent with what we have found for the power spectra in Section~\ref{sect: powerspectrum}.

\begin{figure*}
\begin{center}
\advance\leftskip -0.8cm
\scalebox{0.45}{
\includegraphics[angle=0]{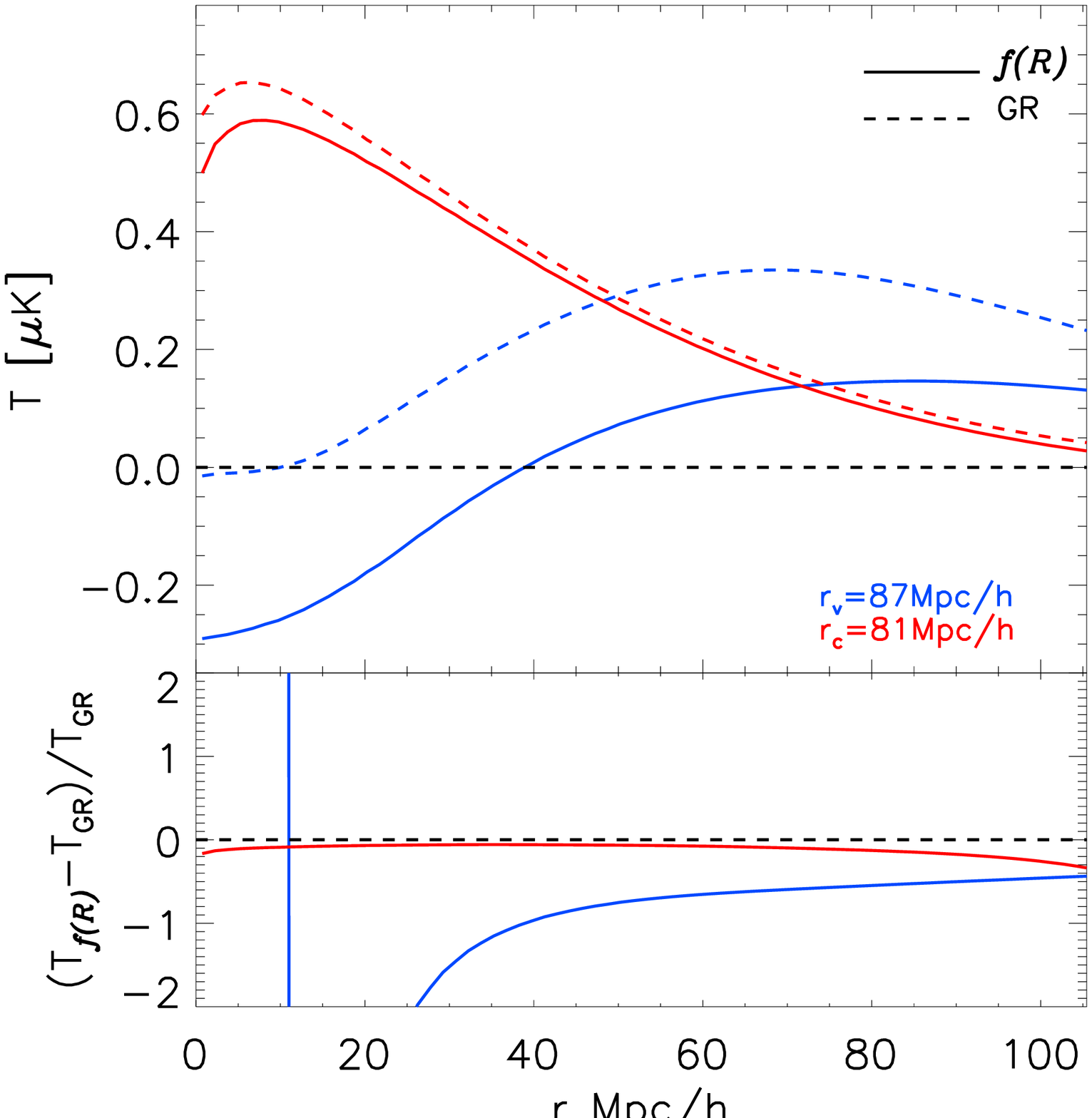}
\includegraphics[angle=0]{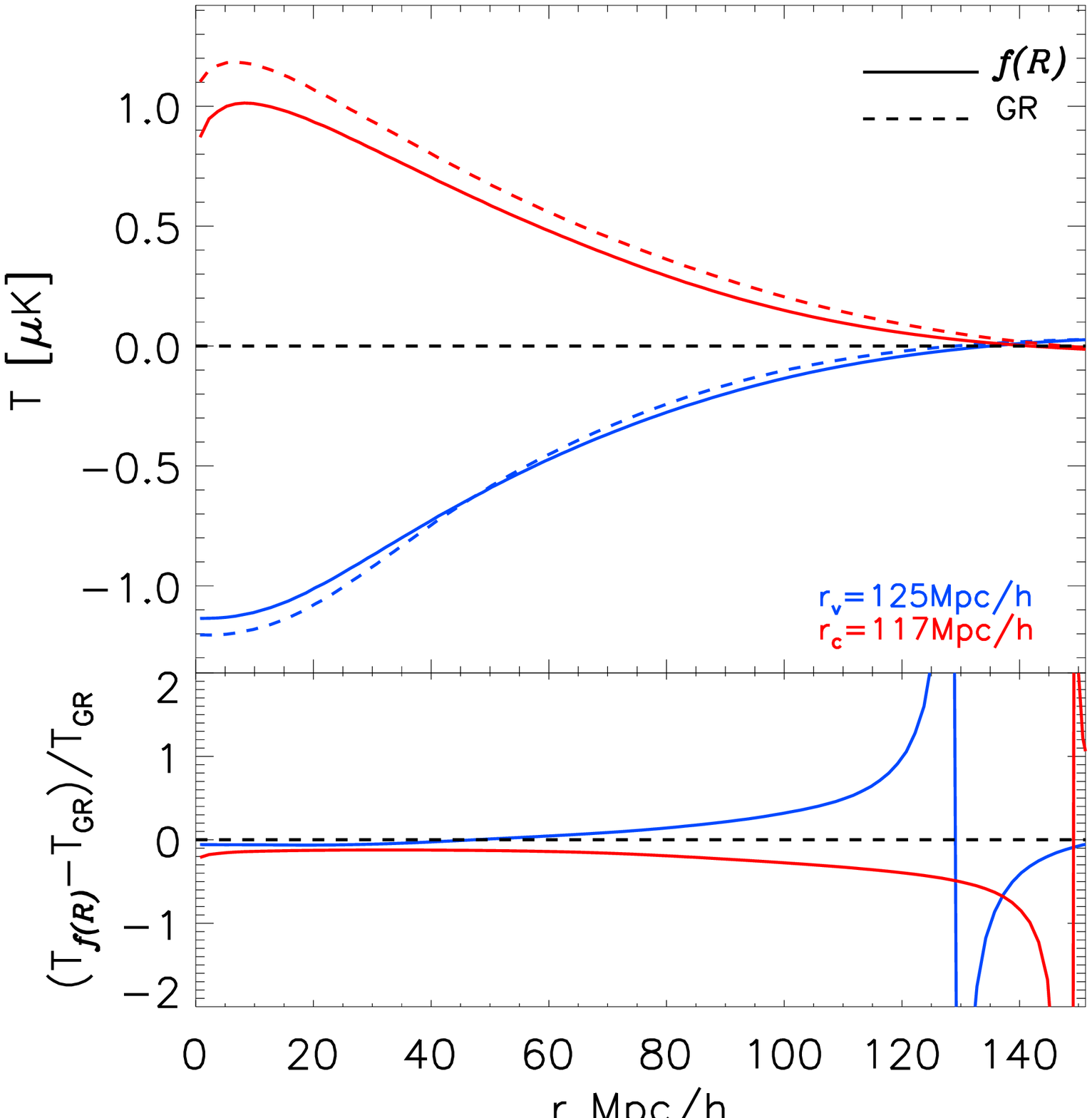}}
\caption{The cumulative ISW temperature profiles from Fig. \ref{StackingClusters} \& Fig. \ref{StackingVoids}. 
blue and red lines are for cold and hot spots respectively. Solid lines shows results from $f(R)$ and 
dashed lines are for GR. The lower panel shows the fractional difference between $f(R)$ and GR. }
\label{ProfilesCumulative}
\end{center}
\end{figure*}

The above qualitative results are supported by the profiles of those cold and hot spots shown 
in Fig.$~$\ref{ProfilesCumulative}. At the centres of stacks, differences between 
$f(R)$ gravity and GR are larger for both supervoids and superclusters. For superclusters (indicated by the red lines), 
the ISW temperatures are suppressed at the centre, and the amplitudes are relatively low in $f(R)$ models, 
while for supervoids (blue lines), whether the amplitude of the cold spots in $f(R)$ gravity is larger or smaller 
depends on the sizes of those structures.

\begin{figure*}
\begin{center}
\advance\leftskip -0.8cm
\scalebox{0.45}{
\includegraphics[angle=0]{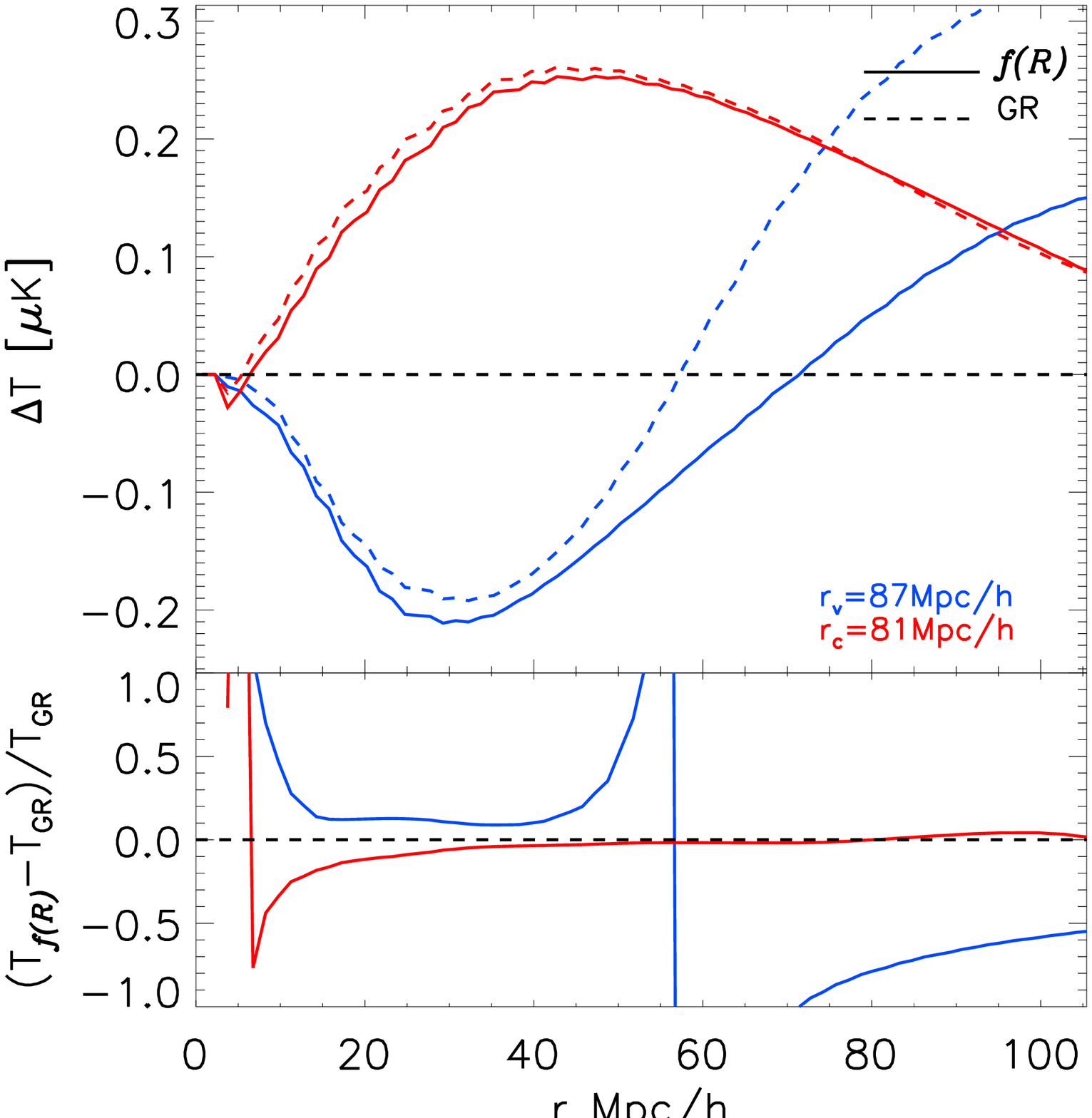}
\includegraphics[angle=0]{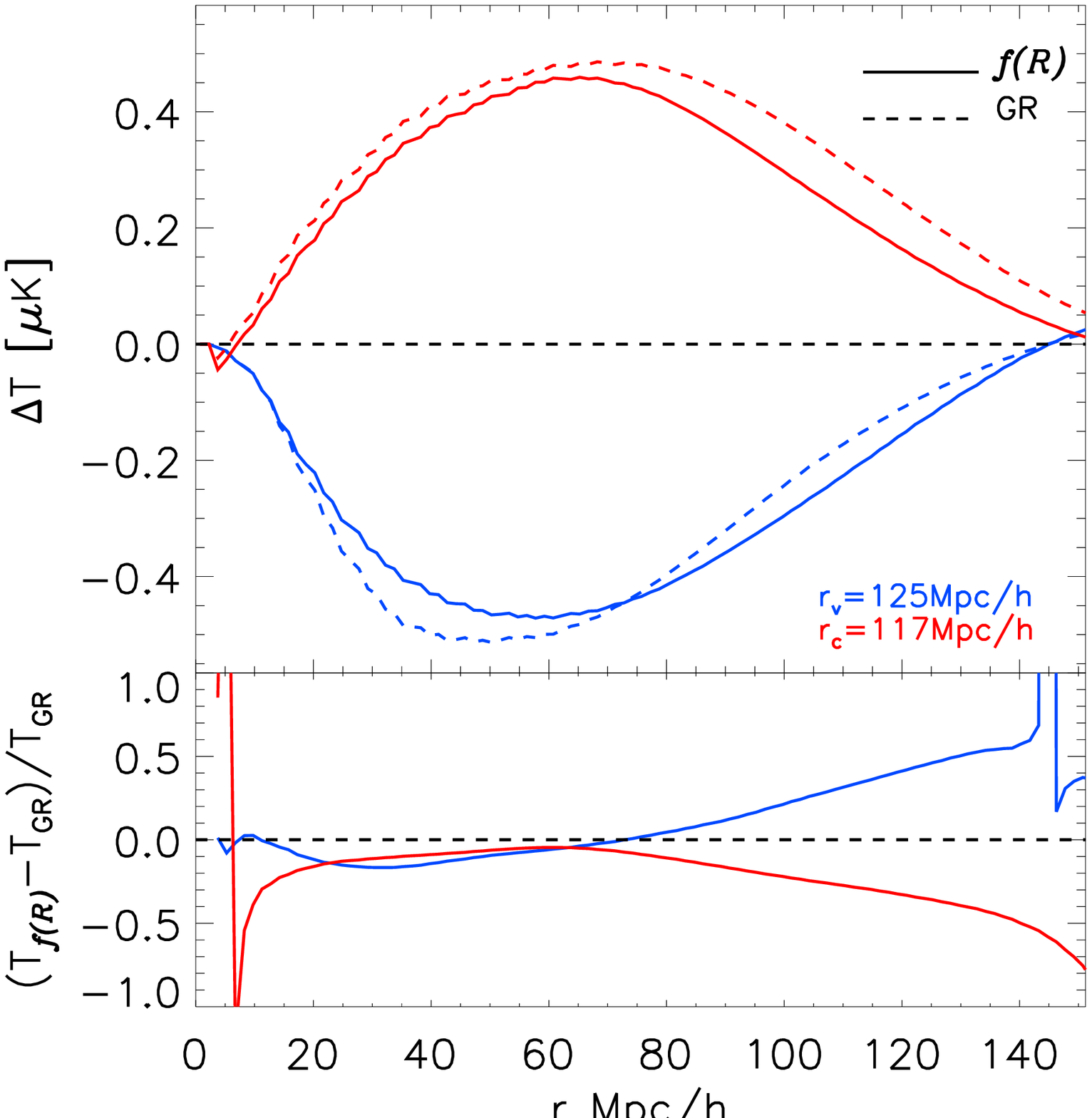}}
\caption{Similar to Fig.$~$\ref{ProfilesCumulative} but showing results from applying compensated Top-Hat filters 
of different sizes \fbtwo{(indicated by $r$ in the $x$-axis)} to the maps shown in Fig. \ref{StackingClusters} \& Fig. \ref{StackingVoids}. 
\fbtwo{These are essentially showing the filtered ISW temperature as a function of filter radius $r$. To plot these lines, we vary the size of 
the compensated Top-Hat filter many times within a large range of scale, and obtain the filtered temperature each time. The $x$-axis, unlike that in Fig. \ref{ProfilesCumulative}, is now indicating the size of 
the filter, rather than the radius of superstructures. }}
\label{ProfilesTopHat}
\end{center}
\end{figure*}

\fb{We note that the stacked ISW signal is strongly affected by cosmic variance. The ISW signal is always dominated by the very large scale perturbation modes, 
i.e. the hot/cold spots corresponding to the superclusters and supervoids are embedded in large scale ISW temperature fluctuations that are much larger than 
the sizes of the superstructures. Due to the differences in structure formation in these two models (even with the same initial conditions in simulations), we 
do not usually find exactly the same superstructures at the same locations between $f(R)$ and GR simulations. When stacking for superstructures of a certain 
range in size, the large scale environments between $f(R)$ and GR models may differ, which will affect the sizes and overall amplitudes of the cold/hot spots. 
This is evident by that fact that the corresponding amplitudes of the blue lines (profiles of the cold spots) in Fig. 4 may be obviously different, but when 
applying the compensated Top-Hat filter, which is efficient in eliminating large-scale perturbation modes and reducing cosmic variance, the differences between 
the filtered profiles are much smaller (see the corresponding blue lines in Fig.$~$\ref{ProfilesTopHat}). Therefore, applying the compensated Top-Hat filter, as also 
to mimic what has been done in observational studies \citep[e.g.][]{Granett2008, Ilic2013, Cai2013}, enable us to see the model differences at the scale of the 
structures that are used for the stack. We find from Fig.$~$\ref{ProfilesTopHat} that the scale where the filtered ISW signal peaks is not the same as that of the 
structures but smaller. It is about 0.5 of the average radius of the structures. This is about the same as found in \citet{Cai2013}. This is unexpected given that 
the selection of superstructures are somewhat different.  
} 

The amplitudes of the cold and hot spots from either of these two models seem relatively small, at the order of 
$1 \mu$K. The level of differences is also minor, i.e. perhaps no more than 20\% for large superstructures. 
The fractional differences for smaller supervoids may be larger, but the amplitudes of them being at the sub-$\mu$K level makes it 
challenging for observation. 

In principle, supervoids and superclusters defined by tracers of dark matter (i.e. halos in our case or galaxies from observations) may not 
resemble those in the dark matter. The stacked ISW signal may have been diluted because of the noise in the simulated catalogues. 
The differences between models may also be affected. However, given the challenges of having the real density field from 
observations (except for lensing), using tracers of dark matter to define superstructures as we have done is perhaps more 
realistic and useful to make comparison with observations.

\fbtwo{The above results uses superstructures with radius greater than 70 Mpc/$h$ because the number of smaller supervoids
passing the 3$\sigma$ selection is too small that the resulting stacked ISW signal is still dominated by cosmic variance. To overcome this, 
we have also repeat our analysis without applying the 3$\sigma$ selection so that the number of small supervoids is increased 
(but we have also certainly introduced noise). In this case, superstructures with the radius between 40 to 70 Mpc/$h$ can be investigated. 
We find that the same qualitative results for small supervoids and superclusters are confirmed, that the ISW cold spot is colder and 
larger in $f(R)$ than in GR and the ISW hot spot is still less hot in $f(R)$.} 

We also find that stacking relatively small supervoids yields ISW hot spot that are much greater than the size of supervoids. 
This may indicate that those small supervoids are more likely to live in large-scale over-dense environments, which are 
undergoing contraction. This cautions using ISW stacks of relatively small supervoids, \fb{see a more detail study in} \citep{Cai2013}. Careful study of  
the void environment may be necessary for understanding the stacking signal.

We have also repeated the same stacking using ISW maps with relatively small $k$-modes removed, i.e. 
$k<0.03h/$Mpc. This may enable us to observe the non-linear ISW feature more clearly by further reducing the noise from 
large-scale $k$-modes, but there is risk that the largest cold and hot spots may be affected. Indeed, all 
our results shown in the above are confirmed. We also find the cold and hot spots are smaller than those in 
Fig.~\ref{StackingClusters} \& \ref{StackingVoids}. This indicates that they are affected by the removal of the small $k$ modes. 

\begin{figure*}
\begin{center}
\advance\leftskip -1.2cm
\scalebox{0.7}{
\includegraphics[angle=0]{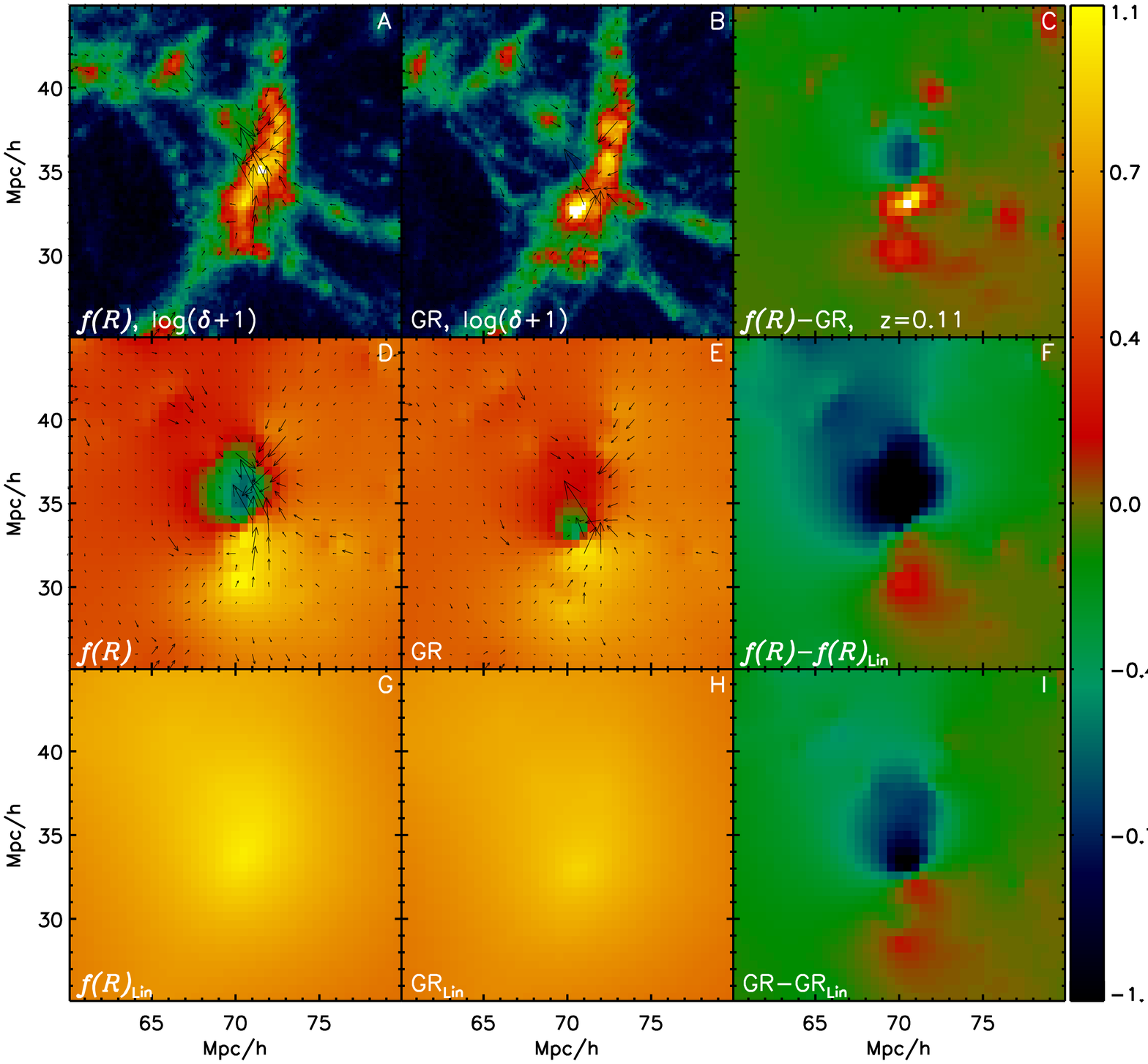}}
\caption{The projected density and ISW temperature maps from a volume of $30 \times 30 \times 30$Mpc/$h$ selected from our simulations. 
The projection is along one principal axis of the simulation box. Arrows are the corresponding mass-weighted velocity of the dark matter. 
The simulations of $f(R)$ and GR models start from the same initial conditions. A \& B -- log density in $f(R)$ and GR simulations; 
D \& E -- the projected linear plus non-linear ISW temperature maps in $f(R)$ \& GR; G \& H -- the linear ISW effect 
calculated from the real matter density field; C -- the ISWRS 
temperature difference between $f(R)$ and GR (D-E); F -- the ISWRS temperature difference between ISWRS 
versus ISW in $f(R)$ (D-G);  I -- the ISWRS temperature difference between ISWRS versus ISW in GR (E-H). }
\label{MovingHalo}
\end{center}
\end{figure*}

\section{fast moving superclusters}
\label{sect: movingcluster}

In this section, we present an example of an ISW map in the deeply non-linear regime where
the difference between $f(R)$ gravity and GR may be most significant: high speed moving (super)clusters.

The transverse motions of massive lumps of dark matter induces time variations of the 
potential for light-rays along different paths. Because of the transverse motion, the potential at the leading part of the moving cluster 
is deepening, hence will cool down CMB photons travelling through it along the line of sight. For the same reason, 
the potential at the trailing part is becoming shallower and CMB photons will be heated up. This leads to dipole features in the 
ISW temperature map, i.e. the leading part of the moving object is cold and the trailing part is hot. \citep[e.g.][]{Cai2010, Molnar2013}.
This effect can equally be thought of as Ômoving lensesÕ, and has been explored by many others in the literature \citep[e.g.][] 
{Birkinshaw1983, Gurvits86, Tuluie95,Tuluie96, Aghanim98,Molnar00, Aso02, Molnar2003, Rubino-Martin04,Cooray05,Maturi07b, Cai2010, Molnar2013}. 
This is also the root cause of the suppression of ISW hot spot in over-dense regions because the 
convergence flow is like a ring of dipoles converging at the centre. Fig.$~$\ref{MovingHalo} gives an example of a supercluster 
that is moving at high transverse velocity. \mk{Note that a moving, possibly linear-scale supercluster need not be entirely in motion. 
It could just be its center (likely, a cluster) that is in motion.}

Seen from the projected density field shown in 
Fig.$~$\ref{MovingHalo}-A \& Fig.$~$\ref{MovingHalo}-B, 
it is an over-dense region with dark matter merging towards the potential centre along 
filaments. Meanwhile, the dark matter at the very central over-dense region seems to be dragged by structures 
north-west of it, and is moving towards that direction. The (mass weighted) velocity field, as indicated by the arrows, is 
stronger in the $f(R)$ simulation, where some sub-structures have merged into the main supercluster, while they are still relatively separated 
from it in the GR simulation. The stronger velocity field causes a larger non-linear ISW effect in the $f(R)$ simulation, for which the dipole is larger and the amplitude 
of the temperature is higher, as evident by Fig.$~$\ref{MovingHalo}-D \& Fig.$~$\ref{MovingHalo}-E. 
However, if linear approximation is made for calculating the ISW signal, a featureless ISW hot spot is found as in
Fig.$~$\ref{MovingHalo}-G \& Fig.$~$\ref{MovingHalo}-H. This is expected as the supercluster is an over-dense region. 
The relatively large differences between Fig.$~$\ref{MovingHalo}-D \& Fig.$~$\ref{MovingHalo}-G, Fig.$~$\ref{MovingHalo}-E \& 
Fig.$~$\ref{MovingHalo}-H highlight the importance of having the full calculation rather than taking linear approximation.
As shown by Fig.~\ref{MovingHalo}-C, Fig.~\ref{MovingHalo}-F \&  Fig.~\ref{MovingHalo}-I, the differences between $f(R)$ gravity and GR, or 
$f(R)$ gravity and GR versus their linear version are all very significant, easily by more than 100\%. 

The amplitude of the dipole feature is of the order of $\sim 1 \mu$K, which corresponds to $\Delta T/T_{\rm CMB}\sim 10^{-6}$, or 
$0.1$ km/s in terms of the frequency shift of photons. Assuming that noise from the CMB is at the order of 10$\mu$K, to reach 
a 3$\sigma$ detection of the dipole feature, one needs to stack 900 fast moving systems. This may not seem many, 
given that current and future large surveys like DES, SPT, EUCLID \tcm{\citep{DES2005,Carlstrom2011,Amiaux2012}} may find hundreds of thousands of clusters. 
However, one needs to select massive clusters that are moving transversely to the line of sight, 
and stack them at the right orientation. This might be challenging, and imposing some selection will inevitably 
reduce the size of the sample. \fb{One possible way to track the directions of motion of clusters is to look for close pairs of clusters, 
which are likely to be in the process of merging. They are likely to be moving towards each other.}

The detection of the dipole in the CMB may be challenging. 
However, if spectra of distant galaxies are used as the background source instead of the CMB, 
the $\sim0.1$ km/s shift of emission or absorption lines may be detectable in the future high resolution spectrographs. 
The idea is that in the case where the moving (super)cluster as a gravitational lens induces multiple galaxy images 
at the background, there would be relative shifts of photon frequencies (of the order of $0.1$ km/s) among different 
lensed images. This is because the spectra should be exactly the same among different lensed images if there is no other 
effects, while photons from multiple images traversing through different parts of the supercluster, where the time variation 
of the potential is different, induce frequency shifts of different amount. The dipole map 
in Fig.~\ref{MovingHalo}-D \& Fig.~\ref{MovingHalo}-E are essentially the map of frequency shifts. 
Given that the mass distribution of the supercluster can be reconstructed from lensing measurement, the relative shifts of 
emission or absorption lines from multiple lensed images can be used to measure the transverse motion of the supercluster 
\citep{Birkinshaw1983, Molnar2003, Molnar2013}. As noted by \citet{Molnar2013}, the frequency shift of $\sim 0.5$ km/s is within the range of the Atacama 
Large Millimetre/Sub millimetre Array for molecular emission CO(1-0), and are near the resolution limit of the new generation 
high-throughput optical-IR spectrographs. In principle, multiple lensed images of quasars can significantly increase the 
statistics and hence the detectability from having a greater number of absorption lines in each quasar spectrum, \fb{but the absorbers have to be 
close enough to the quasars so that the light paths are still about the same when they are absorbed, and as they propagate further, split into different 
multiple images.} \fb{Note that this method relies on having multiple lensed images, which may have a separation of a few hundred kpc at the best. One can perhaps 
only measure the tangential velocities of the lenses at the similar scales for each individual clusters.}

\section{conclusion and discussion}
\label{sect: conclusion}

Using large suites of N-body simulations, we have explored the physics of linear and non-linear ISW effects in $f(R)$ models 
and made comparison with that of the standard $\Lambda$CDM model. In $f(R)$ models, the enhanced gravity speeds 
up the formation of structure, which counters the effect of cosmic acceleration. This affects the evolution of cosmic potentials 
differently at different scales. 
\\
\\
$\bullet$ On large 
scales, where density perturbations are close to linear and the evolution of potential perturbations is dominated by the cosmic 
acceleration, the enhanced growth rate slows down the decay of cosmic potential perturbations and makes the linear ISW 
effect weaker in $f(R)$ models. It reduces the amplitude of the $\dot\Phi$ power spectrum and the amplitudes of 
ISW cold and hot spots.
\\
\\
$\bullet$ On small scales, non-linearity overcomes cosmic acceleration in determining the 
evolution of cosmic potential perturbations. \tcm{The stronger non-linearity in $f(R)$ models enhances the growth of potential wells over 
that in GR}. The ISW effect in this regime is like an enhanced non-linear version of it in GR. 
The amplitude of the $\dot\Phi$ power spectrum is therefore greater. However, since in GR, the non-linear ISW effect 
in supervoids and superclusters behaves differently, cold spots in $f(R)$ models will be colder while hot spots are less hot 
in this regime. 

In summary, ISW hot spots in $f(R)$ models are always less hot than that in GR regardless of their sizes. 
While cold spots may be colder if they are relatively small, but less cold if they are larger scale. 


When stacking supervoids and superclusters of the radius of $\sim$100 Mpc/$h$ in the SDSS DR6 LRG galaxies samples, 
\citet{Granett2008} have found cold and hot spots with the temperatures of $\sim10 \mu$K viewed with a compensated 
top-hat filters. Such a high temperature, if ISW, is unlikely to occur in a $\Lambda$CDM universe. 
There has been speculation about whether alternative models like modified gravity can generate these 
cold and hot spots more naturally \citep{Clampitt2013}. Seen from our simulations of $f(R)$ gravity, even when choosing an $f(R)$ 
model that differs most from GR, it still seems unlikely that those cold and hot spots 
can be accommodated. Cold spots of $\sim$80Mpc$/h$ in radius can be significantly colder in $f(R)$ models, but the 
overall amplitudes of them are 10 times smaller than observed values. For larger cold spots ($\sim$120 Mpc/$h$), 
or hot spots in general, the amplitudes of them are even less than that in GR, which makes it even more unlikely to be an 
explanation of the observations.

With the relatively small difference between $f(R)$ and GR ISW cold and hot spots, and that the 
measurement is expected to have a low signal to noise ratio, it is unlikely to distinguish these two models by 
stacking superstructures to detect ISW effect. \fb{Alternative probe like CMB lensing may be able to set a tighter constraint 
in $f(R)$ \citep[e.g.][]{Marchini2013a, Marchini2013b}, but careful modeling of the non-linear lensing potential is needed.}

Finally, we note that transverse moving clusters, given their strong non-linearity that can amplify the model difference, 
might be a promising test bed for modified gravity. This novel idea relies on the capability of future spectrographs being able to 
detect relative frequency shifts of spectrum lines at the sub-km/s level.

\section*{Acknowledgments}
YC is supported by the Durham Junior Research Fellowship.
\tcr{BL thanks the supports from the Royal Astronomical Society and Durham University.}
YC BL,SC and CF acknowledge a grant with the RCUK reference ST/F001166/1.
The calculations for this paper were performed on the ICC
Cosmology Machine, which is part of the DiRAC Facility
jointly funded by STFC, the Large Facilities Capital Fund of
BIS, and Durham University. Access to the simulations used in 
this paper can be obtained from the authors. \fb{We thank the referee for very detail and useful comments.} 
\bibliography{ISW_MG}
\bibliographystyle{mn2e}

\end{document}